\documentclass[twocolumn,aps,pre,floatfix]{revtex4-2}
\usepackage{psfrag,epsfig,amsfonts,amssymb,amsmath,wasysym,bm}
\usepackage{bbold}
\usepackage{color}
\usepackage{soul}

\usepackage[inline]{enumitem} 
\usepackage{hyperref}

\newcommand{\ket}[1]{\lvert #1 \rangle} 	
\newcommand{\bra}[1]{\langle #1 \rvert}	
 
\renewcommand{\Re}{\operatorname{Re}}

\newcommand{\kB}{k_{\mathrm B}} 

\renewcommand{\ss}{s}

\newcommand{\RR}{{\mathbb R}}

\newcommand{\NN}{{\mathbb N}}

\newcommand{\tr}{\operatorname{tr}}

\newcommand{\lmat}{\left( \begin{matrix}}	
\newcommand{\rmat}{\end{matrix} \right)}	

\newcommand{\LD}[1]{\textcolor{black}{#1}}
\newcommand{\PR}[1]{\textcolor{black}{#1}}

\begin{document}

\title{Allosteric impurity effects in long spin chains}

\author{Christian Eidecker-Dunkel}
\author{Peter Reimann}
\affiliation{Faculty of Physics, 
Bielefeld University, 
33615 Bielefeld, Germany}
\date{\today}

\begin{abstract}
Allosterism traditionally refers to local changes in an extended object, 
for instance the binding of a ligand to a macromolecule, 
leading to a localized response at some other, possibly quite remote position.
Here, we show that such fascinating effects may already 
occur in very simple and common quantum many-body systems, 
such as an anisotropic Heisenberg spin chain:
Introducing an impurity at one end of a sufficiently long
chain may lead to quite significant changes of the observable behavior 
near the other end, but not in the much larger region in between.
Specifically, spin autocorrelation functions
at thermal equilibrium are found to exhibit a pronounced 
allosterism of this type.
\end{abstract}

\maketitle

\section{Introduction}
\label{s1}

It is commonly taken for granted
that isolated many-body systems with short-range 
interactions satisfy a locality principle of the following kind:
A single defect, impurity, or other type of 
local modification does not lead to significant changes 
of the systems'  thermal equilibrium properties at sufficiently remote places.
The main message of our present work consists
in the discovery that exactly the opposite behavior actually occurs
already for very simple examples 
such as an anisotropic Heisenberg spin chain:
Thermal equilibrium correlation functions near {\em both} ends of the 
chain may exhibit quite substantial changes upon introducing
an impurity at one end.
Moreover, no significant changes of the thermal equilibrium 
properties are observed throughout the rest of the chain.
Since such a phenomenon is
somewhat reminiscent of the ``action at a distance'' effects in the context of allosteric 
biochemical processes,
we will adopt here the same label of ``allosterism''
to our present case.

Closely related systems, but with the impurity being located in the middle of 
a spin chain with open boundary conditions,
have been recently explored 
quite extensively for example in Refs. \cite{san04,tor14,bre20,pan20,san20}.
The main focus in these works is on 
the (non-)integrability, relaxation, and transport properties of
models which, in the absence of the impurity, are integrable,
and therefore do not approach thermal equilibrium 
after a quantum quench
\cite{ued20,mor18,tas16,gog16,dal16,bor16,lan16,nan15}.
Remarkably, when switching on the impurity, those systems 
were numerically found in Refs. \cite{san04,tor14,bre20,pan20,san20}
to become non-integrable and thus to exhibit thermalization after a quench.
More precisely speaking, upon parametrically changing
the impurity strength, a continuous transition was 
numerically observed, which becomes more and more rapid as the 
chain length is increased, implying that in the thermodynamic 
limit an arbitrarily
weak impurity would be sufficient
to instantly turn a non-thermalizing system into 
a thermalizing one
\cite{pan20} (see also 
\cite{tor14,san20}).

It is well-known, yet quite remarkable in view of these 
integrability-breaking effects
of a {\em mid-chain} impurity, that 
the same type of impurity at the {\em end} of the chain
provably preserves the system's integrability 
\cite{skl88,alc87,bei13}.
Here, we will show that even more remarkable
effects, namely allosterism, may be caused by such an end-impurity.

An important difference compared to the previous 
Refs. \cite{san04,tor14,bre20,pan20,san20}
is that we focus on systems which are
at thermal equilibrium from the outset.
On the other hand, our findings are -- similar
to those in Refs. \cite{san04,tor14,bre20,pan20,san20}  --
mainly based on numerical explorations, complemented by some 
analytical insights.

\section{Setting}
\label{s2}

We consider the familiar anisotropic spin-1/2 Heisenberg chain
(XYZ model), exhibiting open boundary conditions and 
a magnetic impurity at the ``left end'',
\begin{eqnarray}
H = 
g \,\ss_1^z
-
\sum_{l=1}^{L-1}
J_x \ss^x_{l+1}\ss^x_l
+J_y \ss^y_{l+1}\ss^y_l
+J_z \ss^z_{l+1}\ss^z_l
\, ,
\label{1}
\end{eqnarray}
where $\ss^{x,y,z}_l$ are spin-1/2 operators at
the lattice sites $l\in\{1,...,L\}$ (lattice constant one), 
and $g$ quantifies the strength of the impurity.

We mostly restrict ourselves to
even $L$ and pairwise 
different $J_{x,y,z}$
(generalizations will be briefly addressed
in Sec. \ref{s4}).
The reason is that under these conditions 
we numerically observed 
that all the subsequently explored Hamiltonians (\ref{1}) 
did not exhibit any degeneracies,
which in turn allows us to
make some interesting analytical predictions.
Hereafter, we just state those 
predictions whenever appropriate, referring to the 
accompanying Supplemental Material \cite{sm}
for their detailed derivation.
We also prove in \cite{sm} that
$H$ necessarily must exhibit 
degeneracies for $g=0$ if $L$ is odd or the 
$J_{x,y,z}$ are not pairwise different,
hence some of our analytical predictions no longer apply.

As announced, the system is assumed to be at 
thermal equilibrium,
described by a canonical ensemble 
$e^{-\beta H}/\tr\{e^{-\beta H}\}$ with temperature
$\beta^{-1}$ (Boltzmann constant $\kB=1$).
Accordingly, thermal expectation values of
an observable $A$ are given by
\begin{eqnarray}
\langle A \rangle_{\!\rm th} := 
\tr\{A \,e^{-\beta H}\}/\tr\{e^{-\beta H}\}
\label{2}
\end{eqnarray}
and dynamic (auto-)correlation functions by
\begin{eqnarray}
C_t(A) := \langle A A(t) \rangle_{\!\rm th} - \langle A \rangle_{\! \rm th}^2
\ ,
\label{3}
\end{eqnarray}
where $A(t):=e^{iHt} A e^{-iHt}$
(Heisenberg picture, $\hbar=1$).
Furthermore, their real (or symmetrized) part
\begin{eqnarray}
C^r_t(A) :=\Re\{C_t(A)\}
\label{4}
\end{eqnarray}
is usually of major interest
\PR{(see also discussion below Eq.~(\ref{7})).}

\section{Results}
\label{s3}

\begin{figure}
\includegraphics[scale=1]{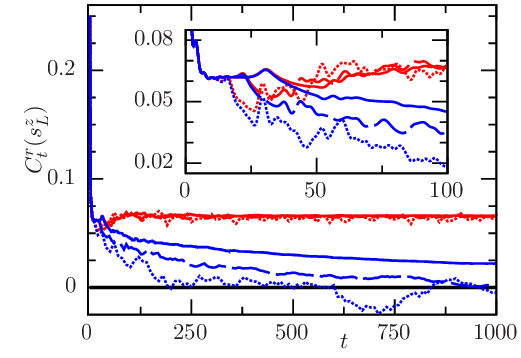}
\caption{Numerically obtained correlation functions $C^r_t(\ss_L^z)$ from 
(\ref{4}) versus time \cite{f1}.
Red: XYZ model (\ref{1}) with $g=0.1$,
$L=20$ (solid), $L=14$ (dashed), $L=10$ (dotted),
$J_x=1$, 
$J_y=1.2$, $J_z=1.5$, 
and $\beta=0.2$. 
Blue: Same, but for $g=0$. 
Inset: Magnification for $t\leq 100$.
}
\label{fig1}
\end{figure}

Figure \ref{fig1} exemplifies the correlations
(\ref{4}) of the magnetization 
$A=\ss_L^z$ at the chain's ``right end'' ($l=L$).
The salient point is that these correlations
exhibit significant differences depending on whether
an impurity at the opposite end ($l=1$) is present 
($g\not=0$, red curves) or not ($g=0$, blue curves). 
The differences become clearly visible for $t \gtrsim 1.5\, L$
(see inset), while all curves \PR{nearly} coincide for 
$t \lesssim 1.5\, L$.
Intuitively, this may be understood as the signature of 
a maximal speed 
at which information about the situation at one end 
can travel to the other end 
\cite{lie72,sto95,bra06,ess16,duv19}.
Likewise, one may understand
why the growth of those differences slows down when $L$ increases
\cite{kem17}.
Yet, they ultimately always approach a sizable, and
asymptotically $L$-independent long-time limit 
(see also the subsequent paragraphs),
i.e., an impurity at one end quite notably affects the
correlation functions at the other end.

\begin{figure}
\includegraphics[scale=1]{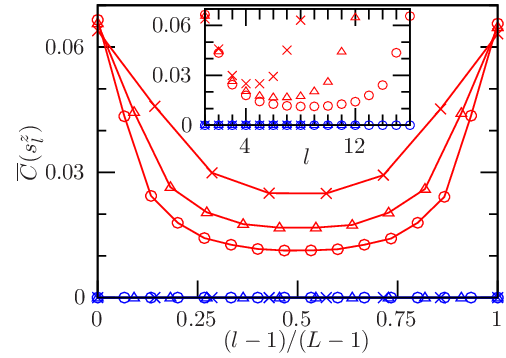}
\caption{Long-time average $\overline{C}(\ss_l^z)$ from 
(\ref{7}) versus chain site $l=1,...,L$ \cite{f1}.
Red: XYZ model (\ref{1}) with $g=0.1$,
$L=16$ (circles), $L=12$ (triangles), $L=8$ (stars),
$J_x=1$, $J_y=1.2$, $J_z=1.5$, and $\beta=0.2$.
Blue: Same, but for $g=0$.
Inset: Raw data. Main plot: Rescaled $x$-axis and 
interpolating lines to guide the eye.
}
\label{fig2}
\end{figure}

Similarly as in Fig. \ref{fig1}, we numerically explored the
correlation functions for various other observables,
most notably or $\ss_l^a$
with $a\in\{x,y,z\}$ and $l\in\{1,...,L\}$.
Again, we found that if some non-negligible difference
in the absence ($g=0$) and in the presence ($g\not=0$)
of the impurity was observable at all,
then this difference predominantly manifested itself 
after sufficiently long times.
Henceforth, we thus restrict ourselves to the
long-time behavior of the correlations $C_t(A)$, 
in particular their long-time average
\begin{eqnarray}
\overline{C}(A):=
\lim_{T\to\infty} \frac{1}{T}\int_0^T dt\,  C_t(A)
\ .
\label{5}
\end{eqnarray}
Indeed, one intuitively expects that,
after initial transients have died out,
the time-dependent correlations $C_t(A)$ 
stay closer and closer to the time-averaged 
value $\overline{C}(A)$ as the system size $L$ increases.
A typical  example is provided by the red curves 
in Fig. \ref{fig1}, 
and further examples can be seen in Figs. S3 and S4 
of the Supplemental Material \cite{sm}.
We also observed this expected long-time behavior 
in all other numerical examples which we explored
with respect to these specific features.
The same has even been shown analytically
under very weak assumptions regarding the spectrum 
of $H$ in Ref. \cite{alh20} (see also \cite{equil}).
Accordingly, we henceforth take it for granted that
$\overline{C}(A)$
faithfully captures the long-time behavior 
of $C_t(A)$.

Denoting by $E_n$ and $|n\rangle$ the eigenvalues and eigenvectors of 
\PR{the Hamiltonian}
$H$ (with $n=1,...,N:=2^L$), 
one readily infers \cite{uhr14,alh20}
from (\ref{2}), (\ref{3}), and (\ref{5})
that
\begin{eqnarray}
\langle A \rangle_{\!\rm th} 
& = & \sum_{n=1}^N p_n \, \langle n|A|n\rangle
\ ,
\label{6}
\\
\overline{C}(A)
& = &
\sum_{n=1}^N p_n\,  \big[\langle n|A|n\rangle - \langle A \rangle_{\!\rm th}\big]^2
\ ,
\label{7}
\end{eqnarray}
where $p_n:=e^{-\beta E_n}/\sum_{m=1}^N e^{-\beta E_m}$ is the population 
of the energy level $|n\rangle$ in the canonical ensemble, and where
-- in case that $H$ exhibits degeneracies 
-- the eigenstates $|n\rangle$ 
must be chosen so that $A$ is diagonal in the 
corresponding eigenspaces of $H$
(see also the Supplemental Material  \cite{sm}).
As an aside, we can infer from (\ref{7}) that
the long-time average in (\ref{5})
must be real and non-negative.
Moreover, it must be equal to the long-time 
average of the real part $C_t^r(A)$ from (\ref{4}).

Assuming $g=0$ (no impurity), 
and exploiting that $H$ exhibits no 
degeneracies (see below Eq. (\ref{1})),
we analytically show in \cite{sm}
that $\langle n|\ss_l^a| n\rangle=0$ for all $n$, $l$, 
and $a\in\{x,y,z\}$.
Hence, the thermal expectation values in (\ref{6}) as well as
the long-time averages in (\ref{7}) must 
vanish for \PR{all} $A=\ss_l^a$.
In particular,
\begin{eqnarray}
\langle \ss_L^z \rangle_{\!\rm th} =0 \ \mbox{and}\
\overline{C}(\ss_L^z) = 0 \  \mbox{if $g=0$}
\, ,
\label{8}
\end{eqnarray}
i.e., all the blue curves in Fig.~\ref{fig1} must
end up by fluctuating around zero.
If $g\not=0$, we furthermore prove in \cite{sm} that 
thermal expectation values and long-time averages still vanish
for all $A=\ss_l^{x,y}$, while $A=\ss_l^z$ must now be evaluated numerically.

Figure \ref{fig2} shows such numerically obtained
long-time averages for $A=\ss_l^z$, 
implying:
(i) As \PR{analytically} predicted, $\overline{C}(\ss_l^z)=0$
for $g=0$ (blue symbols).
(ii) Upon increasing  $L$,
the impurity effects (difference between
blue and red symbols) decrease outside the 
two end regions, while they even slightly 
increase at the two ends.
(iii) The values of $\overline{C}(\ss_l^z)$ and 
$\overline{C}(\ss_{L+1-l}^z)$ are nearly
equal for $g=0.1$ (red symbols).

\begin{figure}
\includegraphics[scale=1]{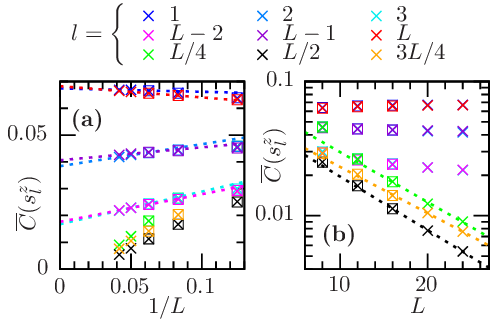}
\caption{(a) Long-time average $\overline{C}(\ss_l^z)$
versus $1/L$ for various $l$-values 
(see legend), employing the same 
model as in Fig.~\ref{fig2} ($g=0.1$).
Crosses:
Numerical results by evaluating (\ref{5}) for large but finite 
$T$ \cite{f1}.
Squares: 
Numerically exact (but also more expensive)
results by evaluating (\ref{7}) via diagonalization of $H$.
Some symbols are (nearly) covered by others.
(b) Same data, but plotted semi-logarithmically
\PR{versus $L$.}
The dotted lines are a guide to the eye,
suggesting for the corresponding $l$-values a
convergence towards a positive large-$L$ 
limit in (a) and an exponential decay towards zero in (b),
see also main text.
}
\label{fig3}
\end{figure}

Concerning (ii), a more detailed finite-size scaling
ana\-lysis is presented in Fig. \ref{fig3}.
Our first remark is that crosses and squares 
have been obtained by means of 
two entirely different numerical methods.
Their close agreement indicates 
that our numerics is trustworthy.
\PR{We also note that the} 
crosses were numerically less expensive,
hence larger $L$ values could be achieved.
Furthermore, the dotted lines in
Fig. \ref{3}(a) indicate that
$\overline{C}(\ss_l^z)$ converges,
for an arbitrary but fixed $l\in\{1,2,3\}$, 
towards a non-vanishing limit when $L\to\infty$,
and likewise when keeping $L-l\in\{0,1,2\}$
fixed.
While these limiting values clearly decrease
with increasing $l\in\{1,2,3\}$ or $L-l\in\{0,1,2\}$,
it nevertheless seems reasonable to expect that 
$\overline{C}(\ss_l^z)$ still asymptotically approaches 
some non-vanishing limit whenever $l$ or $L-l$ is kept 
at an arbitrary but fixed value as $L\to\infty$.
Indeed, the alternative option that the limit 
stays finite up to some maximal distance from the chain ends, 
and then strictly vanishes, appears less reasonable.
On the other hand, for an arbitrary but fixed 
$l/L\in\{1/4,1/2,3/4\}$ the numerical data
in Fig. \ref{fig3}(a) apparently
approach zero faster than $1/L$,
while the dotted lines in Fig.~\ref{3}(b) indicate that they 
asymptotically decrease exponentially with $L$.
Again, it thus seems reasonable to expect such an
exponential decay whenever $l/L$ converges to a limit different 
from zero and unity.
Additional data in support of these predictions
are provided in \cite{sm}.

Altogether, we thus recover the announced allosteric 
impurity effects for long spin chains,
complemented by rather interesting finite-size scaling properties.

\begin{figure}
\includegraphics[scale=1]{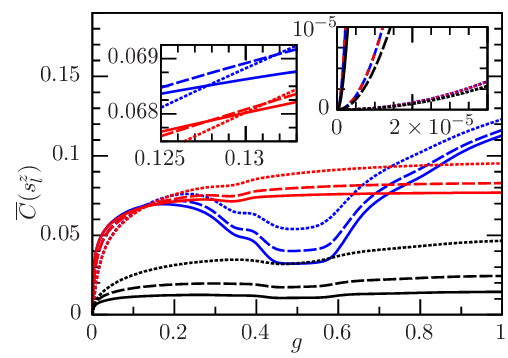}
\caption{Long-time average $\overline{C}(\ss_l^z)$ from 
(\ref{7}) versus impurity strength $g$ \cite{f1}
for $l=1$ (blue), $l=L$ (red), $l=L/2$ (black), 
$L=16$ (solid), $L=12$ (dashed), $L=8$ (dotted),
$J_x=1$, $J_y=1.2$, $J_z=1.5$, and $\beta=0.2$.
Insets: Magnifications near 
$g=0.13$ (left) and $g=0$ (right).
}
\label{fig4}
\end{figure}

\begin{figure}
\includegraphics[scale=1]{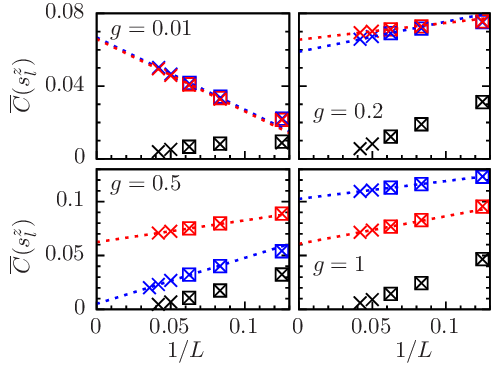}
\caption{Same as in Fig. \ref{fig3}(a)
for $l=1$ (blue), $l=L$ (red), $l=L/2$ (black), 
but now for different coupling strengths 
$g$ (see legends).
For $g=0.5$, an extremely expensive
extra  blue cross at $L=28$ has been 
generated in support of a non-vanishing 
large-$L$ limit.
}
\label{fig5}
\end{figure}

Next we turn to the dependence of the long-time 
correlations $\overline{C}(\ss_l^z)$ on the impurity strength $g$.
Focusing on the ends and the center of the chain, 
i.e., on $l\in\{1,L/2,L\}$, a 
numerical example is depicted in Fig. \ref{fig4}.
Furthermore, a finite-size scaling analysis
analogous to that in Fig. \ref{fig3}(a)
is provided in Fig. \ref{fig5} for a few
representative $g$-values.
The corresponding counterpart of
Fig. \ref{fig3}(b) is available as
Supplemental Material \cite{sm}.
Similarly as before, these numerical
findings indicate that $\overline{C}(\ss_l^z)$
converges to a non-vanishing limit for $l=1$ and $l=L$,
and decays to zero for $l=L/2$.
Furthermore, it seems \PR{again} reasonable to 
expect that the same qualitative
large-$L$ asymptotics will apply to
any given $g>0$, and likewise 
for other $l$-values (see above).

Regarding the issue (iii) from above
\PR{(see paragraph below Eq. (\ref{8})),}
Figs. \ref{fig3}-\ref{fig5} confirm
that $\overline{C}(\ss_l^z)$ and 
$\overline{C}(\ss_{L+1-l}^z)$
are indeed nearly indistinguishable
if (and only if) $g$ is sufficiently small.
Further interesting features of Fig. \ref{fig4}
are the local minima near $g=1/2$ (blue and black) and $g=1/3$ (red),
and the closeness of the crossing points 
near $g=0.13$ (left inset).
Additional details are deferred to the Supplemental Material \cite{sm}
since all these ``extra features'' go beyond 
our actual main objective, namely to demonstrate
the occurrence of allosteric impurity effects {\em per se}.

As already mentioned, for $g=0$ we analytically 
established in \cite{sm} that $\overline{C}(\ss_l^z)=0$.
Yet another analytical prediction in \cite{sm}
is the invariance of $\overline{C}(\ss_l^z)$
under a sign change of $g$. This justifies our restriction to 
$g\geq 0$ in the numerics, and suggests
that $\overline{C}(\ss_l^z)\sim g^2$
for asymptotically small $g$.
The right inset of Fig.~\ref{fig4} confirms this
quadratic asymptotics, and indicates that
the curvature $\partial^2\overline{C}(\ss_l^z)/\partial g^2 |_{g=0}$
actually diverges as $L\to\infty$.
Overall, the behavior for $g\to 0$ and $L\to\infty$
is thus quite intriguing, and in fact somewhat reminiscent of the
integrability-breaking effects of an impurity in the middle of the chain
(see \PR{second paragraph in Sec. \ref{s1}}).

\section{Generalizations and discussion}
\label{s4}

Our findings for more general model parameters
are summarized in the following items (a) to (e),
postponing their detailed analytical and
numerical substantiation to 
the \PR{accompanying}
Supplemental Material \cite{sm}:
(a) If $L$ is odd or the $J_{x,y,z}$ are not pairwise different
(see below Eq. (\ref{1})), then our analytical result 
$\overline{C}(\ss_l^z)=0$ for $g=0$ is no longer valid. 
Accordingly, the behavior of  $\overline{C}(\ss_l^z)$ 
for $g=0$, henceforth abbreviated as
$\overline{C}_{\! 0}(\ss_l^z)$, must be numerically 
explored, and the characteristic signatures of
allosterism are now captured by
$\Delta \overline{C}(\ss_l^z):= \overline{C}(\ss_l^z)- \overline{C}_{\! 0}(\ss_l^z)$.
We numerically found that
$\Delta \overline{C}(\ss_l^z)$
still behaves qualitatively similarly as
in Figs.~\ref{fig2}-\ref{fig5}, i.e., the
system again exhibits the same 
allosteric impurity effects as before.
(b)~We also observed a qualitatively similar
behavior for various other $\beta$-values
in (\ref{2}).
(c) The same applies to other values of $J_{x,y,z}$
in (\ref{1}) at least within the realm 
$J_z>J_{x,y}\geq 0$ 
(\PR{and} excepting $J_x=J_y=0)$.
Quite surprisingly, however, the allosteric effects
are found to disappear when $J_x\geq J_{y,z}\geq 0$ or 
$J_y\geq J_{x,z}\geq 0$.
\PR{(d)~It 
is sufficient to focus on non-negative
interactions $J_{x,y,z}$ in Eq.~(\ref{1})
since the behavior in all other cases can be
inferred by symmetry arguments.
(e) Instead of the canonical ensemble in Eq.~(\ref{2}), one may as well employ
a microcanonical ensemble,
i.e., our allosteric effects exhibit the usual equivalence
of ensemble properties \cite{equiv}.}

Figure \ref{fig4} and its discussion, as well as the above
mentioned observation (c) indicate that 
figuring out the basic physical mechanism behind our
allosteric effects represents a very challenging task.
This is further corroborated by the
following two emarks:
(i)~As far as ``ordinary''  thermal expectation values (\ref{2})
are concerned, we never found
any noteworthy impact of the impurity
on the system's equilibrium properties
sufficiently far away from 
that impurity.
In particular, for the observable $A=\ss_L^z$
we thus can conclude with (\ref{8}) that
$\langle \ss_L^z \rangle_{\!\rm th} =0$
is still fulfilled in very good approximation even
if $g\not=0$, and provided $L$ is sufficiently large.
In other words, our present allosteric effects only
manifest themselves in the correlation functions
from (\ref{3}), not in the expectation values from
(\ref{2}), the key signature being
a non-vanishing thermodynamic limit of 
$\overline{C}(\ss_L^z)$ for $g\not=0$.
(ii)~Generally speaking, 
such a non-vanishing thermodynamic limit 
of $\overline{C}(A)$ is \PR{already in itself 
a quite exceptional situation \cite{f5}.}
The only previous examples known to us are
a few, rather special spin-chain and central spin models,
see Refs. \cite{uhr14,sto92,sto00,kem17,mac18,gri19}
and further references therein.
Most notably, XX-chains with open boundary conditions and
local impurities \PR{of} a different 
type than in (\ref{1}) have been numerically and 
analytically explored in Ref. \cite{sto00},
explaining some of the therein observed
impurity effects in terms of localized single-particle 
``boundary modes'' after mapping the model by means of a 
Jordan-Wigner transformation to a formally equivalent system of 
non-interacting fermions.
However, we found that our present allosterism does
{\em not} occur in those \PR{XX-chain} models from \cite{sto00}
(see also item (c) above),
and that the analytical methods from \cite{sto92,sto00,uhr14}
cannot be adapted to explain our present allosteric 
effects.
In particular, the boundary modes from Ref. \cite{sto00}
become meaningless since our present Hamiltonians (\ref{1}) can no 
longer be mapped to a model of non-interacting 
fermions.
We also remark that a similar observation
as in (c) above has been previously reported in 
Sec. 2.4 of Ref. \cite{sto92} for an XY-model 
in the infinite temperature limit, and most importantly,  
{\em without} any impurity \cite{f8}.

On the other hand, many essential features 
of the blue curves in Fig. \ref{fig1} can be analytically understood \cite{kem17}
by means of so-called edge zero modes \cite{fen16}.
In particular, these insights 
are in agreement with our numerical observation that the blue and red 
curves in Fig.~\ref{fig1} nearly
coincide for $t \lesssim 1.5\, L$.
Focusing on the thermodynamic limit $L\to\infty$ is therefore
useless for the exploration of our present allosteric 
effects, as done, for instance, 
in the analytical investigations in Refs. \cite{gri19,nic22}
of certain impurity effects in the XXZ
model.

Incidentally, taking for granted the 
so-called eigenstate thermalization hypothesis (ETH) \cite{dal16} and 
the usual equivalence of ensembles \cite{equiv}, 
one can conclude that the long-time average 
in Eq. (\ref{7}) must asymptotically vanish for large $L$,
and that our present allosteric effects can thus be ruled out.
In fact, the same 
conclusion can already be deduced \cite{alh20} from a
considerably weaker version of this standard 
ETH \cite{dal16}.
While the standard ETH is a still unproven conjecture
concerning non-integrable models,
this weaker version of the ETH is provably fulfilled by
a very large class of translationally invariant models 
(integrable or not) with short-range interactions \cite{weth}.
Overall, an indispensable (but not sufficient)
prerequisite for the 
appearance of our present allosteric effects thus
seems to be a violation of both the standard and the weak 
versions of the ETH, 
which in turn require that the system must be
integrable \cite{dal16} and not translationally 
invariant \cite{alh20,weth}, respectively.

Altogether, it seems reasonable to suspect that our 
allosteric impurity effects might be somehow related to the 
above mentioned concepts of boundary or edge zero modes, 
and to the (weak) ETH, but we are not aware of any previously established 
analytical tools or intuitive arguments which would admit some seizable
further progress along these lines.

\section{Conlcusions}
\label{s5}

Our main result consists in the discovery of allosteric
impurity effects in long spin chains.
These effects seem to us quite remarkable in themselves,
and comparable findings in such relatively simple many-body
systems with short-range interactions have 
to our knowledge never been observed before.
A very interesting next step will be to explore the 
behavior in response to time-dependent
changes of the end-impurity strength \PR{$g$ in Eq. (\ref{1})}.
In the not unlikely case that a notable response 
will again be detectable at the other chain end, 
but not outside the two end regions (for sufficiently long chains), 
this may open up a conceptually new way of secure 
quantum communication.

Superficially, our present allosteric effects might seem 
to be at least conceptually somehow related to
the celebrated non-locality property of quantum mechanics
(Einstein-Podolsky-Rosen paradox),
or to the so-called cluster decomposition principle \cite{cdp},
but a closer look reveals that both of them are in fact
not very helpful for a better understanding of our present 
case \cite{f6}.
Likewise, our findings are somewhat reminiscent of 
allosteric biochemical processes,
while the underlying physical mechanism
and the way in which the effect actually 
manifests itself are clearly very different.

Another main message of our paper is that
the occurrence (or not), as well as the quantitative 
details of our allosteric effects 
depends in a very subtle manner on the various 
model parameters. 
For instance, the relative magnitude of the
three interactions $J_{x,y,z}$ in Eq. (\ref{1})
seems to play a decisive role.
Accordingly,
an analytical or intuitive explanation
of our numerical observations amounts to a quite 
challenging open problem for future \PR{studies}.

\vspace*{0.8cm}
\begin{acknowledgments}
This work was supported by the 
Deutsche Forschungsgemeinschaft (DFG, German Research Foundation)
within the Research Unit FOR 2692
under Grant No. 355031190,
and
by the Paderborn Center for Parallel 
Computing (PC$^2$) within the project 
HPC-PRF-UBI2. 
\end{acknowledgments}


\newpage

\clearpage
\newpage

\onecolumngrid
\begin{center}
\vspace*{0.5cm}
{\bf\large SUPPLEMENTAL MATERIAL}
\\[1cm]
\end{center}
\twocolumngrid
\setcounter{equation}{0}
\setcounter{figure}{0}

\renewcommand*{\citenumfont}[1]{S#1}

\setcounter{section}{0}

\setcounter{subsection}{0}

\setcounter{figure}{0}
\renewcommand{\thefigure}{S\arabic{figure}}

\setcounter{equation}{0}
\renewcommand{\theequation}{S\arabic{equation}}

\setcounter{table}{0}
\renewcommand{\thetable}{S\arabic{table}}

\renewcommand*{\bibnumfmt}[1]{[S#1]}

Section \ref{sec1} provides the derivation of various analytical predictions -- 
mostly regarding the quantities 
\LD{$\langle n|A| n\rangle$ and  $C_t(A)$
-- which were stated throughout the main paper.
Of particular interest will be single-spin observables 
of the form $A=\ss_l^a$, but also a large class of more 
general $A$ will be covered.}

Section \ref{sec2} contains 
the additional numerical examples and remarks,
as announced throughout the main paper.

\section{Analytical Predictions}
\label{sec1}

As in the main paper, we consider XYZ model Hamiltonians
\begin{eqnarray}
H = 
g \,\ss_1^z
-
\sum_{l=1}^{L-1}
J_x \ss^x_{l+1}\ss^x_l
+J_y \ss^y_{l+1}\ss^y_l
+J_z \ss^z_{l+1}\ss^z_l
\, , \ \
\label{s1}
\end{eqnarray}
with an impurity at site $l=1$ and 
spin-1/2 operators $\ss^{a}_l$,
where $l\in\{1,...,L\}$ and $a\in\{x,y,z\}$.
For the time being, the couplings $g$ and $J_a$ 
may still be arbitrary.

Denoting the eigenvalues and eigenvectors of 
$H$ as $E_n$ and $|n\rangle$ (with $n=1,...,N:=2^L$),
the thermal expectation value of an observable $A$,
\begin{eqnarray}
\langle A \rangle_{\!\rm th} := 
\tr\{A \,e^{-\beta H}\}/\tr\{e^{-\beta H}\}
\ ,
\label{s2}
\end{eqnarray}
can be rewritten as 
\begin{eqnarray}
\langle A \rangle_{\!\rm th}
& = &
\sum_{n=1}^N p_n\, \langle n|A|n\rangle
\ ,
\label{s3}
\\
p_n & := & \mbox{$e^{-\beta E_n}/\sum_{m=1}^Ne^{-\beta E_m}$}
\ ,
\label{s4}
\end{eqnarray}
and its dynamic correlation function
\begin{eqnarray}
C_t(A) 
& := & 
\langle A A(t) \rangle_{\!\rm th} 
- \langle A \rangle_{\! \rm th}^2
\nonumber
\\
& = &
\langle A\, e^{iHt}\!Ae^{-iHt} \rangle_{\!\rm th} 
- \langle A \rangle_{\! \rm th}^2
\label{s5}
\end{eqnarray}
(Heisenberg picture in units with $\hbar=1$) as
\begin{eqnarray}
C_t(A) := 
\sum_{m,n=1}^N p_n\, |\langle m|A|n\rangle|^2 e^{i(E_m-E_n)t}
- \langle A \rangle_{\! \rm th}^2
\ . \ \ \ 
\label{s6}
\end{eqnarray}
One thus can conclude that the time-averaged correlations
\begin{eqnarray}
\overline{C}(A):=
\lim_{T\to\infty} \frac{1}{T}\int_0^T dt\,  C_t(A)
\label{s7}
\end{eqnarray}
can be rewritten 
in the form
\begin{eqnarray}
\overline{C}(A)
& = &
\sum_{n=1}^N p_n\,  [\langle n|A|n\rangle]^2
-
\Big[\sum_{n=1}^N p_n\,  \langle n|A|n\rangle \Big]^2
\nonumber
\\
& = &
\sum_{n=1}^N p_n\,  \big[\langle n|A|n\rangle - \langle A \rangle_{\!\rm th}\big]^2
\ ,
\label{s8}
\end{eqnarray}
where, in case of degeneracies, the eigenstates $|n\rangle$ 
must be chosen so that $A$ is diagonal in the 
corresponding eigenspaces of $H$.
[This guarantees that $\langle m|A|n\rangle=0$ for all 
$m\not=n$ with $E_m=E_n$ in (\ref{s6}).]

\subsection{Basic symmetries}
\label{sec11}

Recalling that the operators $2\ss^a_l$ can be identified with
Pauli matrices (in units with $\hbar=1$),
one readily verifies that those $2\ss_l^{a}$ are
unitary and Hermitian operators, and that
$\ss_l^x \ss_l^z=-\ss_l^z \ss_l^x$.
It follows that 
also the so-called {\em spin-flip operator}
\begin{eqnarray}
U_{\! z} := \prod_{l=1}^L 2 \ss^z_l
\label{s9}
\end{eqnarray}
is a 
unitary and Hermitian operator, satisfying
$U_{\! z}=U_{\! z}^\dagger=U_{\! z}^{-1}$, 
and that
\begin{eqnarray}
U_{\! z} \ss_l^x =-\ss_l^xU_{\! z}
\label{s10}
\end{eqnarray}
for  any given $l\in\{1,...,L\}$.
Likewise, one finds that
\begin{eqnarray}
U_{\! z} \ss_l^y & = & -\ss_l^yU_{\! z}
\ ,
\label{s11}
\\
U_{\! z} \ss_l^z & = & \ss_l^zU_{\! z}
\label{s12}
\end{eqnarray}
for all $l\in\{1,...,L\}$.

Let us now assume that $B$ is an arbitrary product
of factors of the form $\ss_l^a$,
where the indices $a$ and $l$ may or may not
be different for each factor.
Let us moreover assume that  the number of factors 
with the property $a\in\{x,y\}$ is even.
For instance, each 
term on the right-hand side of
(\ref{s1}) is of this form.
It readily follows that $U_{\! z}B=B\,U_{\! z}$,
and hence
\begin{eqnarray}
U_{\! z}H=H\,U_{\! z}
\ ,
\label{s13}
\end{eqnarray}
often referred to as {\em spin-flip or $Z_2$ symmetry}
of $H$.
Likewise, if the number of factors 
with the property $a\in\{x,y\}$ is odd, 
then $U_zB=-BU_z$. In particular,
\begin{eqnarray}
U_{\! z} \ss^{a}_l=- \ss^{a}_l U_{\! z} \ \mbox{for $a\in\{x,y\}$}
\label{s14}
\end{eqnarray}
and any $l\in\{1,...,L\}$.

Since $H$ and $U_{\! z}$ commute (see (\ref{s13})), there exists a common
set of eigenvectors $|n\rangle$,
and since $U_{\! z}$ is unitary, all its eigenvalues are of unit modulus,
implying that
$\langle n|U_{\! z}^\dagger AU_{\! z}|n\rangle=\langle n|A|n\rangle$
for any Hermitian operator $A$.
On the other hand, for the specific operators
$A=\ss^a_l$ with $a\in\{x,y\}$ 
we can conclude from (\ref{s14}) that
$\langle n|U_{\! z}^\dagger AU_{\! z}|n\rangle=- \langle n|A|n\rangle$.
Altogether, we thus obtain
\begin{eqnarray}
\langle n| \ss^{a}_l|n\rangle = 0\ \mbox{for $a\in\{x,y\}$}
\label{s15}
\end{eqnarray} 
and any $l\in\{1,...,L\}$.

Assuming that $g=0$ in (\ref{s1}), and generalizing the definition (\ref{s9}) to
\begin{eqnarray}
U_{\! a} := \prod_{l=1}^L 2\ss^a_l
\ ,
\label{s16}
\end{eqnarray}
one finds, similarly as in (\ref{s13}), that $U_{\! a}H=H\,U_{\! a}$ for any $a\in\{x,y,z\}$.
Assuming furthermore that $L$ is even, the operator 
$U_{\! x}$ contains an even number of factors of the form $\ss_l^x$, 
hence it must commute with $U_{\! z}$ (see above (\ref{s13})).
Analogously,  $U_aU_b=U_bU_a$ for all $a,b\in\{x,y,z\}$.
Similarly as above Eq. (\ref{s15}), one thus can conclude that
there must exist a common set of eigenvectors 
$|n\rangle$ for $H$, $U_x$, $U_y$, and $U_z$
with the property
\begin{eqnarray}
\langle n| \ss^{a}_l|n\rangle = 0  \ \mbox{for $a\in\{x,y,z\}$}
\label{s17}
\end{eqnarray} 
and any $l\in\{1,...,L\}$ {\em provided $L$ is even and $g=0$.} 

\subsection{Degenerate and non-degenerate Hamiltonians}
\label{sec12}

{\em If the Hamiltonian $H$ 
in (\ref{s1}) does not exhibit any degeneracies} then
its eigenvectors $|n\rangle$ are unique (up to irrelevant complex 
phases and permutations of $n$),
hence the extra condition below (\ref{s8}) is trivially fulfilled
(for any Hermitian $A$).
Focusing on such Hamiltonians $H$
\LD{and on $A=s_l^a$ with $a\in\{x,y\}$,}
we thus can conclude that (\ref{s8}) and (\ref{s15}) 
apply simultaneously,
\LD{implying}
\begin{eqnarray}
\overline{C}(\ss_l^a) = 0\ \mbox{for $a\in\{x,y\}$}
\label{s18}
\end{eqnarray}
and any $l\in\{1,...,L\}$.
Similarly, one concludes from
(\ref{s8}) and (\ref{s17}) that
\begin{eqnarray}
\overline{C}(\ss_l^a) = 0\ \mbox{for $a\in\{x,y,z\}$}
\label{s19}
\end{eqnarray}
and any $l\in\{1,...,L\}$ {\em provided $L$ is even and $g=0$.} 

On the other hand, {\em if $H$ exhibits at least one degeneracy},
the extra condition below (\ref{s8}) may in general
be violated by the specific basis $|n\rangle$ appearing 
in (\ref{s15}) and (\ref{s17}).
Hence, our present symmetry considerations
do not admit any relevant conclusions with respect to 
the actual key quantity (\ref{s7}) of the main paper.
In particular, the relations (\ref{s18}) and (\ref{s19}) 
may possibly no longer apply, and are indeed
numerically observed to be violated in general
(see also Sec. \ref{sec2} below).

Next we analytically demonstrate the existence of
degeneracies in two important cases:

First, we consider {\em cases with odd $L$ and $g=0$}
in (\ref{s1}).
As said above Eq.~(\ref{s15}), there exists a common
set of eigenvectors $|n\rangle$ of $H$ and $U_{\! z}$.
The corresponding eigenvalues are denoted as
$E_n$ and $\lambda_n$, respectively.
Since $H$ and $U_{\! z}$ are Hermitian, all 
$E_n$ and $\lambda_n$
must be real, and since $U_{\! z}$ is unitary, all
$\lambda_n$ must be of unit modulus.
Focusing on an arbitrary but fixed $|n\rangle$, 
let us define $|\tilde n\rangle:=U_{\! x}|n\rangle$.
Exploiting $U_{\! x}H=HU_{\! x}$ (see below Eq. (\ref{s16})),
it readily follows that 
both $|n\rangle$ and $|\tilde n\rangle$
are eigenvectors of $H$ with the same 
eigenvalue $E_n$.
On the other hand, since $L$ is odd and $g=0$, one finds similarly 
as above Eqs. (\ref{s13}) and below (\ref{s16}) that $U_{\! x}U_{\! z}=-U_{\! z}U_{\! x}$,
implying that $U_{\! z}|\tilde n\rangle=-\lambda_n|\tilde n\rangle$.
Recalling that $U_{\! z}| n\rangle=\lambda_n|n\rangle$ and
that $\lambda_n$ is of unit modulus it follows that
$|n\rangle$ and $|\tilde n\rangle$ must be linearly independent.
In conclusion, {\em for odd $L$ and $g=0$ all eigenvalues of $H$
must be at least two-fold degenerate}.

Second, we again consider {\em cases with 
$g=0$, but now, we moreover
require that $J_x=J_y$} in (\ref{s1}), i.e., we are
dealing with the so-called XXZ model.
It readily follows that the magnetization
$S_{\! z}:=\sum_{l=1}^L \ss_l^z$
commutes with $H$.
Similarly as before, $H$ and $S_{\! z}$ thus exhibit a common
set of eigenvectors $|n\rangle$ with corresponding 
(real) eigenvalues $E_n$ and $\mu_n$, respectively.
Defining again $|\tilde n\rangle:=U_{\! x}|n\rangle$ and
exploiting $U_{\! x}H=HU_{\! x}$ (see below Eq. (\ref{s16})),
both $|n\rangle$ and $|\tilde n\rangle$
must be eigenvectors of $H$ with the same 
eigenvalue $E_n$.
Similarly as in (\ref{s14}), one furthermore finds that
$\ss_l^zU_{\! x}=-U_{\! x}\ss_l^z$, implying
$S_{\! z}U_{\! x}=-U_{\! x}S_{\! z}$ and thus
$S_{\! z}|\tilde n\rangle = -\mu_n|\tilde n\rangle$.
As before, we can conclude that {\em all eigenvalues of $H$
must be at least twofold degenerate with the possible
exception of those with $\mu_n=0$}
(corresponding to the so-called zero magnetization subsector, 
which only exists for even $L$).

For symmetry reasons, the same conclusions
apply whenever {\em $g=0$
and the couplings 
$J_{x,y,z}$ are not pairwise different}.

Generally speaking, the existence of degeneracies can often be 
rigorously deduced from certain symmetries, as exemplified above, 
or from some other {\em a priori} reasons.
On the other hand, to rigorously prove the 
{\em non}-existence of degeneracies
is usually very difficult or even impossible. 
In particular, the absence of any symmetry 
or other {\em a priori} reasons for degeneracies
still does not exclude the existence
of ``accidental 
degeneracies'' (which then are, however, 
often expected or observed to be
structurally 
unstable and thus restricted to a set of model parameter
values of measure zero).
This very difficult task 
to analytically demonstrate
the absence of degeneracies 
goes beyond the scope of our present work.
On the other hand (and as said in the main paper), we 
numerically observed that our above conditions seem to be
not only sufficient but (generically) also necessary 
for the occurrence of degeneracies 
(at least for all examples we actually examined).

On the basis of this numerical evidence we thus henceforth
take it for granted that, generically, the relation {\em (\ref{s18}) 
applies whenever $g\not=0$}.
Likewise, {\em (\ref{s19}) applies in all cases with  $g=0$, even $L$, 
and pairwise different $J_{x,y,z}$}.
\\[1cm]

\vspace*{-1cm}
\subsection{Further symmetry considerations}
\label{sec13}

Incidentally, for $g=0$ the Hamiltonian (\ref{s1}) also satisfies 
the well-known, so-called parity or reflection symmetry \cite{joe13}.
In the absence of degeneracies it follows that
the eigenstates $|n\rangle$ must be at the same time eigenstates 
of the parity operator.
However, this insight is of limited value for our present purposes.
[One only recovers the obvious fact that in the absence of
the impurity ($g=0$) all thermal equilibrium properties
are invariant under $l\mapsto L+1-l$.]
On the other hand, in the above mentioned cases where
the occurrence of (at least two-fold) degeneracies can be 
shown analytically, the parity symmetry may give rise to
more than two-fold degeneracies.

Next, we turn to a generalization of the basic symmetry 
considerations from Sec. \ref{sec11}.
Similarly as above (\ref{s13}), let us denote by $A^+_a$
an arbitrary product of factors of the form $\ss_l^b$,
where the indices $b$ and $l$ may or may not
be different for each factor, but such that
the number of factors with the property 
$b\not=a$ is even.
Moreover, the symbol $A^+_a$ may also represent an
arbitrary linear combination of such operators.
Finally, the symbol $A^+$ represents any linear
combination of such operators $A^+_a$ with arbitrary
$a\in\{x,y,z\}$.
Likewise, the symbols $A_a^-$ refers to odd 
instead of even numbers with the property $b\not=a$,
and analogously for $A^-$.
Similarly as in (\ref{s15}) one then can conclude that
\begin{eqnarray}
\langle n| A^-_z|n\rangle = 0
\label{s20}
\end{eqnarray} 
for any such operator $A_z^-$.
Under the same precondition as in (\ref{s18}), 
which in turn 
applies whenever
$g\not=0$ (see end of previous subsection), 
it follows that
\begin{eqnarray}
\overline{C}(A_z^-) = 0
\ .
\label{s21}
\end{eqnarray}
In the same vein, for $g=0$ and even $L$ the generalization
of (\ref{s17}) takes the form
\begin{eqnarray}
\langle n| A^-|n\rangle = 0
\label{s22}
\end{eqnarray} 
for any $A^-$ as specified above.
Under the additional condition of pairwise 
different couplings $J_{x,y,z}$ one similarly can conclude that
the generalization of (\ref{s19}) 
takes the form
\begin{eqnarray}
\overline{C}(A^-) = 0
\label{s23}
\end{eqnarray}
for any $A^-$.

Note that any observable $A$ can be rewritten
in the form $A^+ + A^-$.
The set of observables covered by (\ref{s22}) 
and (\ref{s23}) is thus already quite large.

\subsection{Symmetries of thermal expectation values and correlation functions 
under a sign change of  $g$}
\label{sec15}

Let us consider Hamiltonians of the general structure
\begin{eqnarray}
H=g H_x^- +H_x^+
\ ,
\label{s24}
\end{eqnarray}
where $H_x^\pm$ are of the form as specified above (\ref{s20}).
A particular example is our original model from (\ref{s1}).
Our main goal is a comparison of 
\LD{such a Hamiltonian (\ref{s24})}
and its
counterpart
\begin{eqnarray}
\tilde H := - g H_x^- +H_x^+
\ .
\label{s25}
\end{eqnarray}
In other words, we are interested in the consequences
of a sign change of $g$ in (\ref{s1}), or, more generally, in (\ref{s24}).
Moreover, we will focus on observables 
$A_x^\pm$ as specified above (\ref{s20}).
Particularly interesting examples of the type $A_x^-$ 
are
\begin{eqnarray}
A_x^-=\ss_l^z
\ .
\label{s26}
\end{eqnarray}

Similarly as above (\ref{s14}), one readily verifies that
\begin{eqnarray}
U_{\! x}A^\pm_x=\pm A_x^\pm U_{\! x}
\ .
\label{s27}
\end{eqnarray}
Together with (\ref{s24}) and (\ref{s25}) we thus obtain
\begin{eqnarray}
\tilde H U_{\! x}=U_{\! x} H
\ .
\label{s28}
\end{eqnarray}
Recalling that $U_{\! x}=U_{\! x}^\dagger$ (since $U_{\! x}$ is simultaneously unitary and Hermitian), 
it is thus natural to generally define
\begin{eqnarray}
\tilde A:= U_{\! x} A U_{\! x}
\label{s29}
\end{eqnarray}
for any given operator $A$.
Exploiting (\ref{s27}) we can conclude that
\begin{eqnarray}
\tilde A_x^\pm=\pm A_x^\pm
\ .
\label{s30}
\end{eqnarray}

As before, we denote by $|n\rangle$ and $E_n$ the eigenvectors and 
eigenvalues of $H$, and we define
\begin{eqnarray}
|\tilde n\rangle := U_{\! x} |n\rangle
\ .
\label{s31}
\end{eqnarray}
Exploiting (\ref{s28}), one readily infers that
$\tilde H|\tilde n\rangle = E_n|\tilde n\rangle$,
i.e., the eigenvectors of $\tilde H$ are related to those of $H$
via (\ref{s31}), and the corresponding eigenvalues are identical
for $\tilde H$ and $H$.
Note that if $H$ exhibits degeneracies, these conclusions 
apply to any choice of the basis $|n\rangle$.

Similarly as for the thermal expectation values 
of a system with Hamiltonian $H$ in (\ref{s2}),
let us abbreviate those of a system with Hamiltonian 
$\tilde H$ as
\begin{eqnarray}
{\langle A \rangle}_{\!\widetilde{\rm th}} := 
\tr\{A \,e^{-\beta \tilde H}\}/\tr\{e^{-\beta \tilde H}\}
\ .
\label{s32}
\end{eqnarray}
Together with Eqs. (\ref{s3}), (\ref{s4}), and
the discussion around (\ref{s31})
we can conclude that
\begin{eqnarray}
{\langle A \rangle}_{\!\widetilde{\rm th}} 
& = & \sum_{n=1}^N p_n \langle \tilde n|A|\tilde n\rangle
\ .
\label{s33}
\end{eqnarray}
Exploiting (\ref{s29}) and (\ref{s31}) we thus obtain
\begin{eqnarray}
{\langle A \rangle}_{\!\widetilde{\rm th}} 
= \sum_{n=1}^N p_n \langle n|\tilde A| n\rangle
\ .
\label{s34}
\end{eqnarray}
Together with (\ref{s30}) and (\ref{s3}) we can conclude that
\begin{eqnarray}
\langle A^\pm_x \rangle_{\!\widetilde{\rm th}} 
= \pm \langle A^\pm_x \rangle_{\!\rm th}
\ .
\label{s35}
\end{eqnarray}

In other words, upon switching the sign of $g$ in (\ref{s1}) (or more generally in (\ref{s24})),
thermal expectation values of observables of the type $A_x^+$ remain unchanged,
while thermal expectation values of observables of the type $A_x^-$ simply change their sign.
Particular examples of the latter type are the spin operators $s_l^z$ in (\ref{s26}).
Further examples will be provided in Eq. (\ref{s38}) below.

In the same vein, the correlation functions (\ref{s5}) can 
be rewritten in the form (\ref{s6}) for a system with 
Hamiltonian $H$,  and (by similar arguments as in the 
derivation of (\ref{s34})) as
\begin{eqnarray}
\tilde C_t(A) := \sum_{m,n=1}^N p_n\, |\langle m|\tilde A |n\rangle|^2 e^{i(E_m-E_n)t}
-
{\langle A \rangle}_{\!\widetilde{\rm th}}^2
\ \ \ \ \ \ \ 
\label{s36}
\end{eqnarray}
for a system with Hamiltonian $\tilde H$.
Again by means of almost the same steps as before, 
this finally yields the result
\begin{eqnarray}
\tilde C_t(A^\pm_x) =  C_t(A^\pm_x)
\ .
\label{s37}
\end{eqnarray}

In other words, we arrive at our main result that
correlation functions of arbitrary observables of the type $A_x^\pm$ are
invariant under a sign change of $g$ in (\ref{s1}) (or, more generally, in (\ref{s24})).
The same invariance is inherited by the time-averaged
correlation functions (\ref{s7}).

Note that any observable $A$ of the form
\begin{eqnarray}
A=\prod_{k=1}^K \ss_{l_k}^{a_k}
\vspace{1cm}
\label{s38}
\end{eqnarray}
\\[-0.1cm]
with arbitrary $a_k\in\{x,y,z\}$, $l_k\in\{1,...,L\}$, and $K\in\NN$ 
is either of the form $A^+_x$ or of the form $A^-_x$.
For any such observable $A$, the correlation functions $C_t(A)$
are thus invariant under a sign change of $g$,
while the thermal expectation values are either invariant
or inverted (see below (\ref{s35})).

Analogous conclusions as in (\ref{s35}) and (\ref{s37}) are readily 
recovered also for observables of the type $A^\pm_y$.

\subsection{Symmetries under sign changes of two $J_a$}
\label{sec16}

Let us define, similarly as in (\ref{s16}), for any 
given $a \in \left\lbrace x,y,z\right\rbrace$ 
the even-site spin-flip operator 
\begin{eqnarray}
U^e_{a}:=\prod_{{\rm even}\, l} 2s_l^{a}
\ ,
\label{s39}
\end{eqnarray}
where the product runs over all even $l\in\{1,...,L\}$.
Similarly as in Sec.~\ref{sec11}, one can conclude that
\begin{eqnarray}
U^e_{a}s^{a}_l
=s^{a}_lU^e_{a}
\label{s40}
\end{eqnarray}
for all $l\in\{1,...,L\}$ and $a \in \left\lbrace x,y,z\right\rbrace$.
It follows that
\begin{eqnarray}
U^e_{a}s^{a}_l s^{a}_{l+1}&=s^{a}_l s^{a}_{l+1}U^e_{a}
\label{s41}
\end{eqnarray}
for all $l\in\{1,...,L-1\}$ and 
$a \in \left\lbrace x,y,z\right\rbrace$.
Analogously, one finds for arbitrary 
$a,b \in \left\lbrace x,y,z\right\rbrace$ with $a\neq b$ that
\begin{eqnarray}
U^e_{a}s^{b}_l
& = & 
-s^{b}_l U^e_{a} 
\ \ \mbox{for even $l\in\{1, ... ,L\}$,}
\label{s42}
\\
U^e_{a}s^{b}_l
& = & 
s^{b}_lU^e_{a}
\ \ \mbox{for odd $l\in\{1, ... ,L\}$,}
\label{s43}
\\
U^e_{a}s^{b}_l s^{b}_{l+1}
& = & 
-s^{b}_l s^{b}_{l+1}U^e_{a}
\ \ \mbox{for all $l\in\{1, ... ,L-1\}$.} \ \ \ \ \ 
\label{s44}
\end{eqnarray}
Generalizations to 
other observables (cf. previous subsections)
and to odd-site spin-flip operators are straightforward 
but not essential in the following.

Denoting by  $H(g,J_x,J_y,J_z)$ the Hamiltonian \eqref{s1} for an arbitrary 
but fixed set of parameters $g,J_x,J_y,J_z$, and exploiting the various above
derived relations, it follows that
\begin{eqnarray}
U^e_x H(g,J_x,J_y,J_z)&=H(g,J_x,-J_y,-J_z)U^e_x
\label{s45}
\ ,
\\
U^e_yH(g,J_x,J_y,J_z)&=H(g,-J_x,J_y,-J_z)U^e_y
\ ,
\label{s46}
\\
U^e_zH(g,J_x,J_y,J_z)&=H(g,-J_x,-J_y,J_z)U^e_z
\ .
\label{s47}
\end{eqnarray}

Returning to the abbreviation $H:=H(g,J_x,J_y,J_z)$, 
denoting by $|n\rangle$ and $E_n$ the eigenvectors 
and eigenvalues of $H$, and defining
\begin{eqnarray}
|\tilde n\rangle & := & U_x^e |n\rangle
\ ,
\label{s48}
\\
\tilde H & := & H(g,J_x,-J_y,-J_z)
\ ,
\label{s49}
\end{eqnarray}
we thus can infer from (\ref{s45}) that
$\tilde H|\tilde n\rangle = E_n|\tilde n\rangle$,
i.e., the eigenvectors of $\tilde H$ are related to those of $H$
via (\ref{s48}), and the corresponding eigenvalues 
are identical for $\tilde H$ and $H$.
Moreover, by exploiting (\ref{s40}) and (\ref{s42}), one can conclude that
\begin{eqnarray}
\bra{\tilde n} s^{a}_l
\ket{\tilde n}=\pm\bra{n}s^{a}_l\ket{n}
\ ,
\label{s50}
\end{eqnarray}
where the minus sign applies if $l$ is even and 
$a\not =x$, and the plus sign otherwise.

By almost the same line of reasoning as in the
previous subsection one now can conclude
that the thermal expectation values for $A=s^a_l$
in (\ref{s2})-(\ref{s4}) are the same for $H$ and $\tilde H$ 
if the plus sign applies in (\ref{s50}), and of 
opposite sign otherwise.
According to (\ref{s49}), 
this is tantamount to 
a simultaneous sign change of $J_y$ and $J_z$.
Likewise, the correlation functions for $A=s^a_z$
in (\ref{s5})-(\ref{s7}) are found to be invariant 
under a simultaneous sign change of $J_y$ and $J_z$.

By exploiting the symmetry (\ref{s46}) or (\ref{s47})
instead of (\ref{s45}), analogous conclusions are
found to apply for a simultaneous sign change 
of $J_x$ and $J_z$, or of $J_x$ and $J_y$,
respectively.

\begin{figure}[b]
\hspace*{-0.9cm}
\includegraphics[scale=0.99]{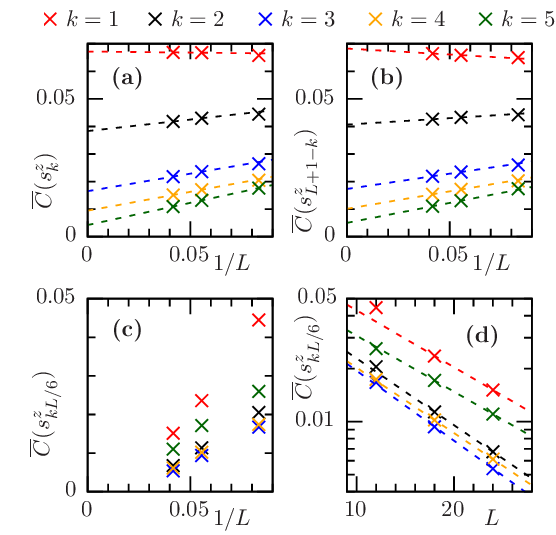}
\caption{
Same as in Fig.\,3 of the main paper, but now for a larger variety of 
different $l$-values.
In more detail: Long-time averages $\overline{C}(\ss_l^z)$ 
by numerically evaluating (\ref{s7}) for large but finite $T$
are depicted for the XYZ model from (\ref{s1})
with $g=0.1$, $J_x=1$, $J_y=1.2$, $J_z=1.5$, and $L\in\{12,18,24\}$,
employing $\beta=0.2$ in (\ref{s2})-(\ref{s5}).
(a) Results for $l=k\in\{1,2,...,5\}$,
corresponding to the 5 leftmost sites of the chain.
(b) Results for $l=L+1-k\in\{L-4,L-3,...,L\}$,
corresponding to the 5 rightmost sites of the chain.
(c) Results for $l=kL/6\in\{L/6,2L/6,...,5L/6\}$,
corresponding to 5 sites in the ``bulk'' of the chain.
[These $l$-values are the reason for our focusing on $L\in\{12,18,24\}$.]
(d) Same data as in (c), but now plotted semi-logarithmically
\LD{versus $L$.}
The dotted lines are a guide to the eye,
suggesting the convergence towards a positive large-$L$ 
limit in (a) and (b), and an exponential decay towards zero in (d).
}
\label{figs1}
\end{figure}

\begin{figure}[b]
\hspace*{-0.3cm}
\includegraphics[scale=1.05]{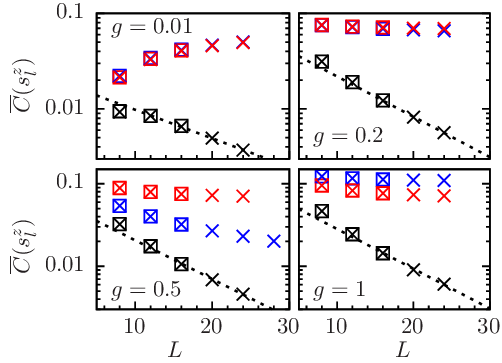}
\caption{
Same data as in Fig.\,5 of the main paper, but now plotted semi-logarithmically
\LD{versus $L$}.
In more detail: Long-time averages $\overline{C}(\ss_l^z)$  from (\ref{s7}) are plotted
versus $L$ for the XYZ model from (\ref{s1})
with $J_x=1$, $J_y=1.2$, $J_z=1.5$, and different $g$-values (see legends),
employing $\beta=0.2$ in (\ref{s2})-(\ref{s5}).
Blue, black, and red symbols refer to $l=1$, $l=L/2$, and $l=L$, respectively.
Crosses: 
Numerical results by evaluating (\ref{s7}) for large but finite $T$.
Squares: 
Numerically exact (but also more expensive)
results by evaluating (\ref{s8}) via 
diagonalization of $H$.
The dotted lines are a guide to the eye,
suggesting an exponentially decaying 
large-$L$ asymptotics of the black symbols.
}
\label{figs2}
\end{figure}

\subsection{Symmetries under a sign changes of one $J_a$}
\label{sec17}

According to (\ref{s1}) and (\ref{s2}),
thermal expectation values of arbitrary observables $A$
are left invariant when simultaneously changing the sign of 
all five model parameters $\beta,g,J_x,J_y,J_z$. 
Likewise, one can infer from (\ref{s5}) 
that dynamical correlation functions exhibit a complex
conjugation under a simultaneous sign change of 
$\beta,g,J_x,J_y,J_z$, while their long-time average in (\ref{s7}) is 
left invariant (note that the right-hand side of (\ref{s8}) 
is purely real).

Focusing on observables of the form $A=s_l^a$, 
dynamical correlation functions were found to be 
invariant under a sign change of $g$ in Sec. \ref{sec15},
and under a simultaneous sign change of two
among the three parameters $J_x$, $J_y$, $J_z$
in Sec. \ref{sec16}.

Upon properly combining all these various possibilities
to change the sign of certain parameters,
we thus can conclude that changing
the sign of $\beta$ and of one among the 
three parameters $J_x$, $J_y$, $J_z$ 
results in a complex conjugation of the 
dynamical correlation functions and leaves
their long-time average invariant, at least for
all observables of the form  $A=s_l^a$.

The corresponding transformation behavior
when simultaneously changing the sign of
two or three of the parameters  $J_x$, $J_y$, $J_z$
follows by sequentially carrying out one sign 
change after the other.

Analogous conclusions can be obtained for 
the thermal expectation values of $s_l^a$, 
but they are of no immediate relevance 
in the main paper.

The overall main result of the last three subsections 
is that one can focus on non-negative $J_a$ and $g$, 
as it is done in the main paper and in the 
subsequent section.

\begin{figure}[t]
\includegraphics[scale=0.95]{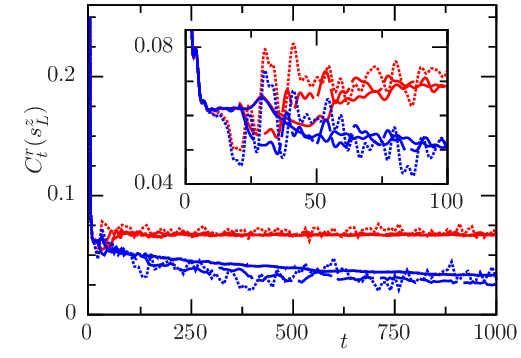}
\caption{Same as in Fig.\,1 of the main paper, but now for odd instead of even
system sizes $L$, namely $L=19$ (solid), $L=13$  (dashed) and $L=9$ (dotted). 
\\[0.3cm]
}
\label{figs3}
\end{figure}

\begin{figure}
\includegraphics[scale=0.95]{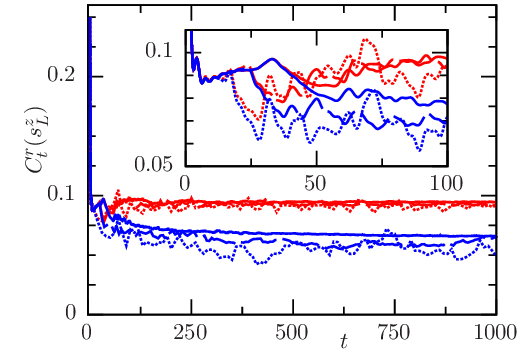}
\caption{Same as in as in Fig.\,1 of the main paper, but now for couplings $J_{x,y,z}$
which are not pairwise different, namely $J_x=J_y=1$, $J_z=1.5$. 
}
\label{figs4}
\end{figure}

\begin{figure}
\hspace*{-0.3cm}
\includegraphics[scale=1.05]{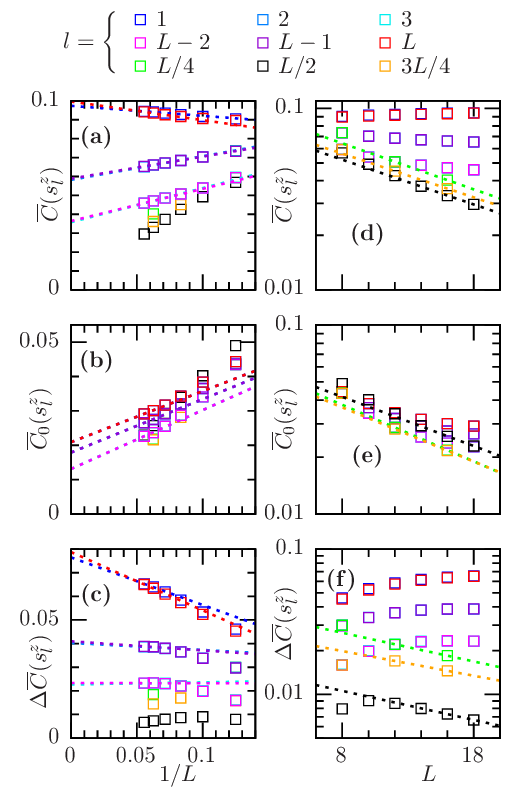}
\caption{(a): Long-time average $\overline{C}(\ss_l^z)$ 
by evaluating (\ref{s8}) via numerically exact diagonalization 
of $H$ versus $1/L$ for various $l$-values (see legend), 
employing the same system as in Fig.\,\ref{figs4}
($g=0.1$ and $J_{x,y,z}$ not pairwise different).
(b) Same, but without impurity ($g=0$), and denoting the
long-time average as $\overline{C}_{\! 0}(\ss_l^z)$.
(c) Difference 
$\Delta \overline{C}(\ss_l^z):= \overline{C}(\ss_l^z)- \overline{C}_{\! 0}(\ss_l^z)$
between the long-time averages in (a) and (b).
(d)-(f): Same data as in (a)-(c) but now plotted semi-logarithmically
versus $L$.
Some symbols are (nearly) covered by others,
and for $L\in\{10,14\}$ there are no green and orange symbols.
The dotted lines are a guide to the eye,
suggesting the convergence towards a
finite large-$L$ limit in (a)-(c), and an exponential decay 
towards zero in (d)-(f).
}
\label{figs5}
\end{figure}

\section{Additional Numerical Examples}
\label{sec2}

Hereafter, our supplemental numerical results are presented
in the order they are announced in the main paper.
Also some additional observations and remarks are
included in the accompanying text.

Fig.\,\ref{figs1} provides 
further numerical evidence for the large-$L$ asymptotics
as detailed in the main paper at the end of the
discussion of Fig.\,3 therein.
As an interesting side remark we observe that
all the slopes (decay rates) in 
\LD{Fig.\,\ref{figs1}\,(d)}
are (nearly) equal, while the offsets (pre-exponential constants) 
increase upon approaching the chain's ends.
This provides some additional insight
how the large-$L$ asymptotics when keeping $l/L$ fixed
matches the asymptotics when keeping $l$ (or $L-l$) fixed.

Fig.\,\ref{figs2} was announced in the discussion of Fig.\,4
in the main paper, plotting once more the same data as in
Fig.\,5 of the main paper, but now on a semi-logarithmic 
scale.
Some further interesting features of Fig.\,4 in the main paper 
will be addressed in the context of Fig.\,\ref{figs7} below.

The remaining Figs. \ref{figs3}-\ref{figs8} illustrate the
items (a)-(c)
in 
\LD{the first paragraph of Sec. IV in}
the main paper.

Specifically, the case of odd system sizes $L$ in item (a) 
is exemplified by Fig.\,\ref{figs3}.
Likewise, Fig.\,\ref{figs4} illustrates item (a) for a case,
where the couplings $J_{x,y,z}$ are not pairwise different
(while $L$ is again even).
Essentially, the findings in both Fig.\,\ref{figs3} and Fig.\,\ref{figs4} are  
qualitatively quite 
similar to the case of even $L$ and pairwise different $J_{x,y,z}$
(see Fig.\,1 in the main paper), except that the blue curves no 
longer approach zero for large times $t$
(see also Fig.\,\ref{figs5} below).
In other words, our analytical prediction 
$\overline{C}(\ss_l^z)=0$ for $g=0$ from (\ref{s19})
is no longer applicable
since (as detailed in Sec. \ref{sec12}) the Hamiltonian
is known to exhibit degeneracies if $L$ is odd
or the $J_{x,y,z}$ are not pairwise different.

Accordingly, if $L$ is odd or the $J_{x,y,z}$ are not pairwise different,
the quantitative value of 
$\overline{C}(\ss_l^z)$ for $g=0$, henceforth abbreviated 
as $\overline{C}_{\! 0}(\ss_l^z)$, 
must be numerically 
determined.
Fig.\,\ref{figs5} exemplifies the finite-size scaling
behavior of the long-time averages 
$\overline{C}(\ss_l^z)$ (for $g=0.1$) and
$\overline{C}_{\! 0}(\ss_l^z)$ (for $g=0$)
in such a case where the $J_{x,y,z}$ are 
not pairwise different.

\begin{figure*}
\includegraphics[scale=1]{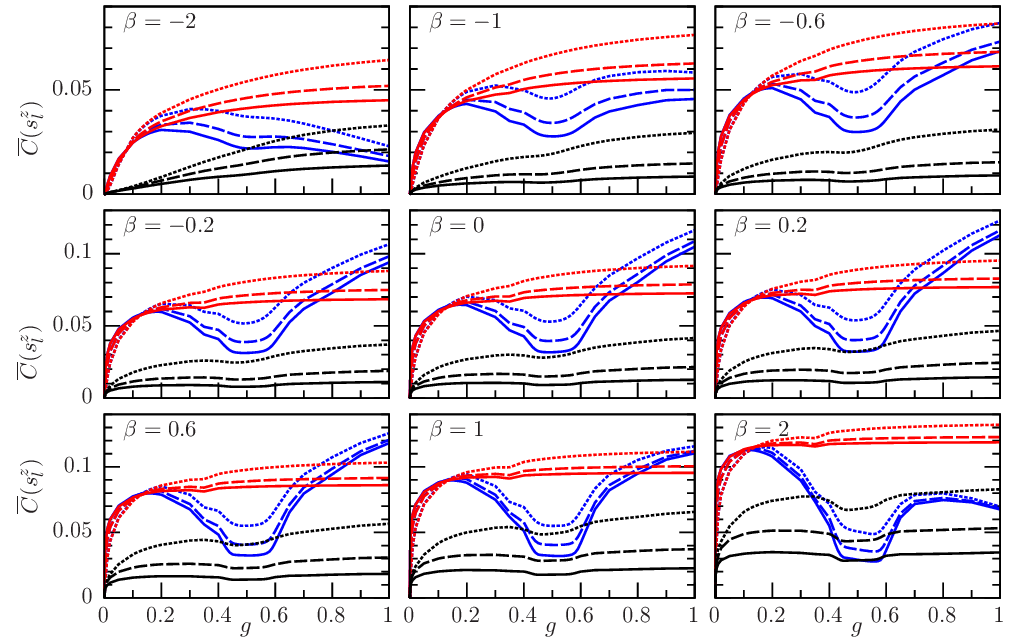}
\caption{Same as in Fig.\,4 of the main paper, but now for various other values of $\beta$ (see legend). 
}
\label{figs7}
\end{figure*}

In view of the quite slowly decaying blue curves in 
Fig.\,\ref{figs4}, the time-averaged correlations in Fig.\,\ref{figs5}
were {\em not} obtained by numerically evaluating (\ref{s7})
for large but finite $T$, but rather by the numerically exact 
but also more expensive method to diagonalize $H$
and then evaluate (\ref{s8}). For this reason, the maximally
achievable $L$-values in Fig.\,\ref{figs5} are smaller than in 
our previous figures of this type.

Also shown in Fig.\,\ref{figs5} is the quantity 
$\Delta \overline{C}(\ss_l^z):= \overline{C}(\ss_l^z)- \overline{C}_{\! 0}(\ss_l^z)$.
As already pointed out in the main paper, this quantity captures the characteristic 
signatures of allosterism in such a case where our ``usual''
analytical prediction $\overline{C}_{\! 0}(\ss_l^z)=0$
no longer applies.
In other words,  Figs.\,\ref{figs5}\,(c) and (d) essentially amount to the
counterparts of Figs.\,3\,(a) and (b) in the main paper, respectively,
and indeed exhibit the same qualitative behavior.

Fig.\,\ref{figs7} serves as a numerical illustration of item (b) in the main paper,
namely the dependence of our findings on the parameter 
$\beta$ in (\ref{s2})-(\ref{s8}).
Note that the spectrum of the Hamiltonian (\ref{s1}) is bounded
for any finite $L$, hence all the quantities in (\ref{s2})-(\ref{s8})
are perfectly well-defined for arbitrary $\beta\in\RR$.
We also recall that the behavior for negative
$\beta$ is of particular interest with respect to the 
symmetry properties under a sign change of one (or three)
of the couplings $J_a$ (see Sec. \ref{sec17}).
We finally remark that $\beta$ represents the inverse temperature
(in units with Boltzmann constant $\kB=1$), and that the
physical relevance and meaning of negative temperatures
are sometimes still considered as not satisfactorily settled issues.
This ongoing debate is only of very little relevance with respect to
our actual main issues and therefore not further pursued.

Overall, the basic qualitative features in Fig.\,\ref{figs7}
change remarkably little upon variation of $\beta$.
Without going into the details we mention that 
much stronger finite $L$-effects than in Fig.\,\ref{figs7} 
are expected and observed to arise for even much larger (positive) $\beta$-values
(low temperatures), caused by two nearly 
degenerate lowest energy eigenvalues, 
which are separated by a gap from  the rest of the spectrum.

Intuitively, one generally expects that the spin 
$s_l^z$ at the site of the impurity ($l=1$) will be 
forced for asymptotically large $g$ to align with 
the impurity in (\ref{s1}) and
hence the correlations in (\ref{s5}) will approach zero.
On the other hand, the correlations away from the impurity
site ($l>1$) are not expected to approach zero but rather
to approach some non-trivial (positive) large-$g$ limit.
Moreover, these effects are expected to require increasingly
large $g$-values upon reducing the value of 
$\beta$ in (\ref{s2})-(\ref{s8})
(and may possibly even disappear for $\beta=0$).
Along these lines, one also
may understand why the blue 
curves start to decrease in 
Fig.\,\ref{figs7} for $\beta=\pm 2$ as $g$ approaches
unity, but not the red and black curves,
and not for the smaller values of $|\beta|$ in 
Fig.\,\ref{figs7}. 

\begin{figure}[b]
\hspace*{-0.3cm}
\includegraphics[scale=1.05]{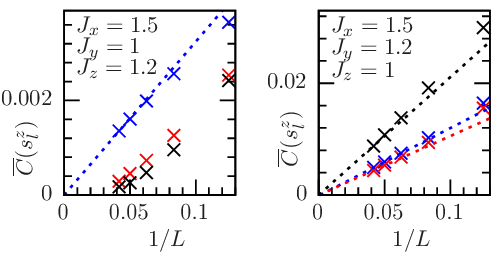}
\caption{
Same as in Fig.\,5 of the main paper, but now 
for other values of the couplings $J_{x,y,z}$ (see legends),
and for $g=0.1$ (see also Fig.\,\ref{figs1}, Fig.\,\ref{figs2}, 
and Fig.\,3 in the main paper).
In particular, the crosses once again represent time-averaged correlations 
$\overline{C}(\ss_l^z)$ for $l=1$ (blue), $l=L$ (red) and $l=L/2$ (black).
}
\label{figs8}
\end{figure}

On the basis of these considerations it also appears 
intuitively quite evident that a non-monotonic dependence 
of $\overline{C}(\ss_l^z)$ on $g$ is, generally speaking, 
nothing unexpected, and likewise for the possibility
that $\overline{C}(\ss_l^z)$ may assume smaller 
values at the location of the impurity ($l=1$) than
at the opposite chain end ($l=L$).
The appearance of the extra (local) minima in Fig.\,\ref{figs7}
is harder to explain by simple arguments, but it seems
not unlikely that somewhat similar basic
mechanisms may be at work.

Finally, Fig.\,\ref{figs8} exemplifies item (c) in the main paper,
namely the disappearance of our allosteric impurity effects
when $J_z$ is no longer the largest among all three (positive)
couplings $J_{x,y,z}$
(regarding negative couplings see Sec. \ref{sec17}).
The salient point in Fig.\,\ref{figs8} is that the correlations 
$\overline{C}(\ss_l^z)$ now indeed asymptotically approach 
zero for large $L$ at both ends as well as in the 
middle of the chain.
Whether $\overline{C}(\ss_l^z)$ decays exponentially with $L$,
or as $1/L$ (or even in some other way) apparently still 
depends on $l$ and on the choice of the couplings 
$J_{x,y,z}$.
Moreover,  the (approximate) equality of 
$\overline{C}(\ss_l^z)$ and $\overline{C}(\ss_{L+1-l}^z)$
(for sufficiently small $g$), which we observed in all previous
examples, seems to be violated in the left plot 
in Fig.\,\ref{figs8}.
A more detailed exploration of these issues will be presented elsewhere.



\begin{thebibliography}{70}
\bibitem{f1}
Numerically, we evaluated  (\ref{3}) 
by utilizing standard dynamical typicality and 
Runge-Kutta propagation methods,
and (\ref{7}) by diagonalization of $H$.
\bibitem{f5}
\PR{For a very large class of translationally invariant models 
it has been rigorously shown in Ref. \cite{alh20}
that $\overline{C}(A)$ must go to zero in the 
thermodynamic limit.
The opposite behavior of  $\overline{C}(A)$ 
in our present model and in the spin-chain and 
central spin models in 
Refs.~\cite{uhr14,sto92,sto00,kem17,mac18,gri19}
does not contradict Ref. \cite{alh20}
since translational invariance is violated
in all these models.}
\bibitem{f8}
\PR{Namely,  setting $J_z=g=0$ in (\ref{1})
and $\beta=0$ in (\ref{2}), the long-time average 
$\overline{C}(\sigma_1^x)$ for $L\to\infty$
was found  in Sec.~2.4 of Ref. \cite{sto92} 
to be positive if $J_x>J_y>0$
and zero if $J_y\geq J_x>0$.}
\bibitem{cdp}
\PR{The cluster decomposition principle means that
the spatial (equal-time) correlations of local operators vanish 
in the limit of large separations between  
them, see e.g. Refs.~\cite{mor18,gog16} 
for recent reviews and  Refs. ~\cite{wic63,wei97} 
for particularly early pertinent works.}
\bibitem{f6}
\PR{Some reasons are: In our case, it is important that the 
system is at thermal equilibrium, exhibits many degrees 
of freedom, and that temporal (rather than spatial)
correlations are considered. On the other hand, 
the fundamental subtleties of the quantum mechanical 
measurement process play no prominent role in our case.}
\end{thebibliography}
\end{document}